\documentclass[letterpaper]{jpconf}
\usepackage{graphicx,psfrag, amssymb}

\usepackage{citesort}

\newcommand{\cmt}[2]{\mbox{\begin{minipage}[t]{#1cm} #2 \end{minipage}}}
\newcommand{\cmb}[2]{\mbox{\begin{minipage}[b]{#1cm} #2 \end{minipage}}}
\newcommand{\M}{\mathbb{M}}
\newcommand{\be}{\begin{displaymath}}
\newcommand{\ee}{\end{displaymath}}
\newcommand{\bne}{\begin{equation}}
\newcommand{\ene}{\end{equation}}
\newcommand{\intersect}{\cap}
\newcommand{\meff}[1]{m_#1^{\mathrm{eff}}}
\newcommand{\past}{\mathrm{past}}

\newcommand{\mod}{\:\mathrm{mod}\:}

\begin{document}
\title{Emergence of spatial structure from causal sets}

\author{D Rideout$^1$ and P Wallden$^2$}

\address{$^1$ Perimeter Institute for Theoretical Physics, 31 Caroline St N, Waterloo, ON N2K~1S5, Canada}
\address{$^2$ Raman Research Institute, Sadashivanagar, Bangalore - 560 080, India}

\ead{$^1$ drideout@perimeterinstitute.ca, $^2$ petros@rri.res.in}

\begin{abstract}
There are numerous indications that a discrete substratum underlies continuum
spacetime.  
Any fundamentally discrete approach to quantum gravity must provide some
prescription for how continuum properties emerge from the underlying
discreteness.  The causal set approach, in which the fundamental relation is
based upon causality, finds it easy to reproduce timelike distances, but has
a more difficult time with spatial distance, due to the unique combination of
Lorentz invariance and discreteness within that approach.  We describe a
method to deduce spatial distances from a causal set.  In addition, we sketch
how one might use an important ingredient in deducing spatial distance, the
`$n$-link', to deduce whether a given causal set is likely to faithfully embed
into a continuum spacetime.
\end{abstract}

\section{Introduction}

\subsection{Fundamental Discreteness}

There are many reasons to believe that some sort of Planck scale discreteness
will be present in any theory of quantum gravity.  Perhaps the most
convincing evidence is the finiteness of black hole entropy, which requires a
cutoff at around the Planck scale.  Thus we expect quantum gravity to be
described as a discrete theory of geometry, in some form or other.  
One can 
regard the continuum as \emph{emerging} from the discrete, in a similar way
in which continuum theories for fluids emerge from the underlying physics of
their discrete molecules.

In the case of fluids, the discrete elements (molecules) naturally live in a
continuum background, and inherit their interrelationships from their
embedding in this background.
The situation for gravity, however, is different.  There it is natural to
expect that the discrete elements do not live in a background medium.
In order to recover geometry, in this absence of a background, one must
impose 
some sort of relationship among the discrete
elements, for example in the form of a graph or binary relation,
which indicates which discrete elements are `nearest neighbors' in
spacetime.  

It is important to note in this vein that nature appears to satisfy local
Lorentz invariance, at least to the degree accessible by current
observations.  Since the intuitive notion of nearest neighbors is naively
frame dependent, a fundamental relation for discrete spacetime must contain
the nearest neighbors for every Lorentz frame.  Thus one expects that the
number of nearest neighbors of any element will be infinite (or if finite
they should generically extend to cosmological scales).  From these heuristic
arguments
one may expect that the discreteness of quantum gravity may be expressible,
in its simplest form, as
some sort of very highly connected graph or binary relation.

As suggested above, it makes sense to guess that a fundamental relation for
discrete quantum gravity would be compatible with the Lorentz symmetry.
Stated the other way around, one might expect that the fundamental relation
is such that it allows the macroscopic Lorentz symmetry to emerge naturally
from the discrete.  There are two possibilities which come to mind, one which
takes a microscopic causal ordering as this relation, and another which takes
spatial nearest neighbors, as given by the invariant spacetime interval
$ds^2$.  
It is not possible to specify both freely because the causal relation alone
is sufficient to recover the metric up to a conformal factor.  Thus given one
relation, it must be possible to deduce the other.

A number of approaches to quantum gravity take the second approach, such as
loop quantum gravity\footnote{Loop quantum gravity does not literally take
  this approach, however the fundamental object with which the theory is
  expressed is purely spatial in character, and the edges of the spin network
  encode (among other things) a binary spatial relation on the vertices.}
 \cite{thiemann,ashtekar,rovelli},
 `quantum graphity' \cite{qgraphity}, and the approach outlined in \cite{nks}.
Quantum graphity, for example,
has as its fundamental discrete
structure a graph which is regarded as a spatial object, which evolves in
time.  It postulates a form for the Hamiltonian which describes its time
evolution, and from this one is able to derive the presence of light cones
and non-trivial causal structure \cite{leib_robinson,nks}.

Here we take the alternate approach, that microscopic causal ordering is fundamental, and show that it
is possible to derive from it a symmetric, spacelike relation, at least for
discrete structures (causal sets) which are well approximated by Minkowski
space.  This approach has the advantage that one can address the question of
how continuum structures emerge from the discrete, without reference to a
particular dynamical law.

The ability to recognize continuum properties from the discrete relation may
be an important stepping stone toward constructing a full theory of quantum
gravity.  In particular, most of our understanding of how to formulate
gravitational dynamics is within the continuum.  It would be helpful to
understand how this carries over to the discrete context.  
This understanding may also provide crucial hints as to a fundamental
origin of black hole entropy and the covariant entropy bound, by providing
clues as to how to count states associated with a spacelike or null hypersurface
\cite{entropy_bound}.

Once we have the ability to deduce spatial distances between discrete
elements in the approximating continuum, it is not difficult to use this to
compose a symmetric, spatial nearest neighbor relation, which can be useful
for recovering topology and geometry of curved spacetime.  In addition this
may allow contact with other approaches to quantum gravity, which hold the
spacelike relation to be fundamental.

We address the question of how to deduce the spatial distances in an
approximating continuum, given only a discrete partial order.
It turns out that, while the recovery of timelike distances from the discrete
causal ordering is relatively straightforward, recovering spacelike distances
is much more difficult, due to the relatively unfamiliar nature of Lorentz
invariant discreteness.

\subsection{Causal Sets}

The causal set approach to quantum gravity is based on two observations.  One
is that the causal structure of a spacetime, namely a list of which events
`can causally influence' which others, is a very rich structure, enough to
reconstruct the conformal metric.  The other is the abundance of evidence
suggesting that some sort of discrete structure underlies continuum
spacetime. \cite{causet_prl}

The intent is to approach `quantization' via histories, i.e.\ to take the
\emph{histories}, rather than states defined on some `spatial hypersurface',
as the fundamental objects of the theory.  The quantum theory can then be
expressed in terms of some appropriate generalization of the Feynman path
integral \cite{qmt,coevent,gqt}.

In the causal sets approach, the histories are taken to be \emph{causal
  sets}, which are countable sets of `atoms of spacetime', endowed with a
partial order relation $\prec$ which is transitive and irreflexive.  For a
causal set $C$, transitivity requires that if $x \prec y$ and $y \prec z$
then $x \prec z$, $\forall x, y, z \in C$.  Irreflexivity is simply the
condition that no element can precede itself, namely $x \not\prec x \; \forall x
\in C$.  To enforce discreteness, one imposes a local finiteness condition,
that every \emph{order interval} $[x,y] = \{z \in C | x \prec z \prec y\}$
has finite cardinality.

The correspondence between the discrete causal set and a continuum spacetime
is described in terms of a \emph{faithful embedding}, which is an order
preserving map $\phi$ from a causal set $C$ to a spacetime $(M,g)$, which has
the property that ``the number of elements mapped into any region of spacetime
volume $V$ is Poisson distributed with mean $V$''.  Here \emph{order
  preserving} means that two events $\phi(x)$ and $\phi(y)$ in the image of
$\phi$ are causally related in the spacetime ($\phi(x) \in J^-(\phi(y)$) iff
the corresponding elements in $C$ are related $x \prec y$.  The Poisson
distribution with mean $V$ assigns a probability of $\frac{V^n e^{-V}}{n!}$
to the event of mapping $n$ elements to the region of volume $V$.

A faithful embedding is easy to realize, one simply selects points in the
target spacetime $(M,g)$ at random by a Poisson process (a `sprinkling'), and then computes a
partial order on them using the causal structure of the spacetime.  This
gives a causal set $C$, along with a faithful embedding $\phi$ of $C$ into
$(M,g)$.  It is useful to make use of the map $\phi$ implicitly when
discussing the causal set, so that one can speak of ``the causet elements in a
region of, or at points in, $(M,g)$'' to refer to those elements which are
mapped to regions of $(M,g)$ by $\phi$.

It is important to keep in mind, however, that it is the discrete causal set
that is fundamental, and the continuum into which it may faithfully embed is
to be regarded only as an approximation to the discrete substructure.  The
faithful embedding is merely a tool with which to discuss how the continuum
arises from the discrete.
It is worth remarking that it may be the case that a
`physical' causal set may only faithfully embed into a continuum spacetime
after some coarse graining, so the physics near the Planck scale need not be
continuum-like.

\begin{figure}[htb]
\begin{center}
\cmt{8.2}{
\psfrag{512 Element Causet Faithfully Embedded into SxI}{}
\psfrag{space}{space}
\psfrag{time}{time}
(a)
\includegraphics[width=8.2cm]{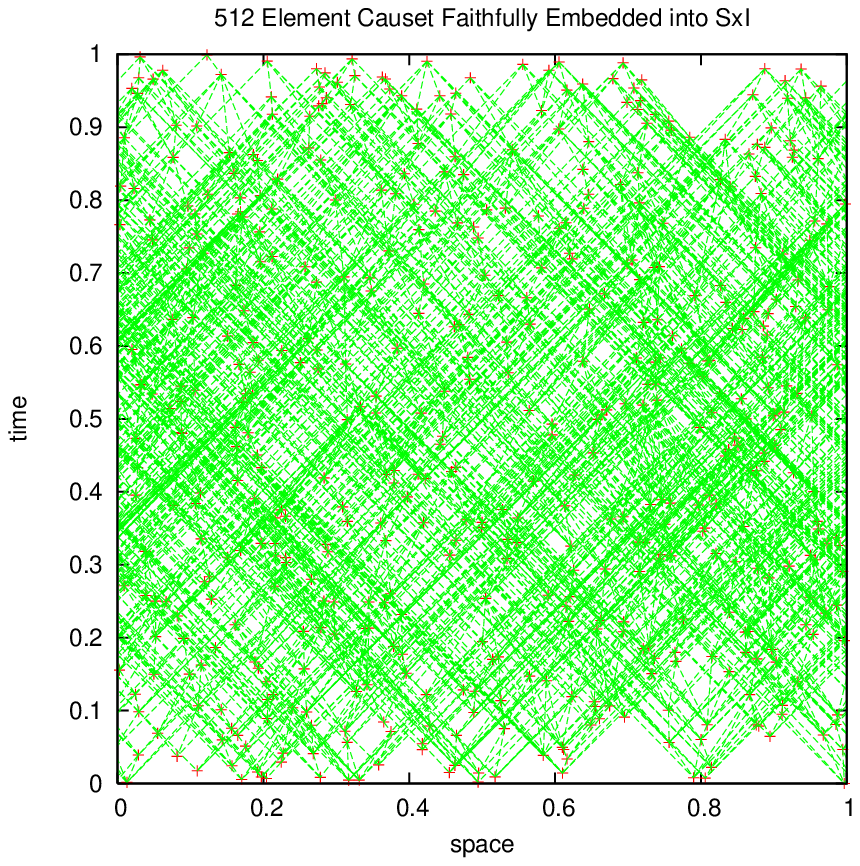}}
\cmt{7.65}{
\psfrag{space}{space}
\psfrag{time}{time}
(b)
\includegraphics[width=7.65cm]{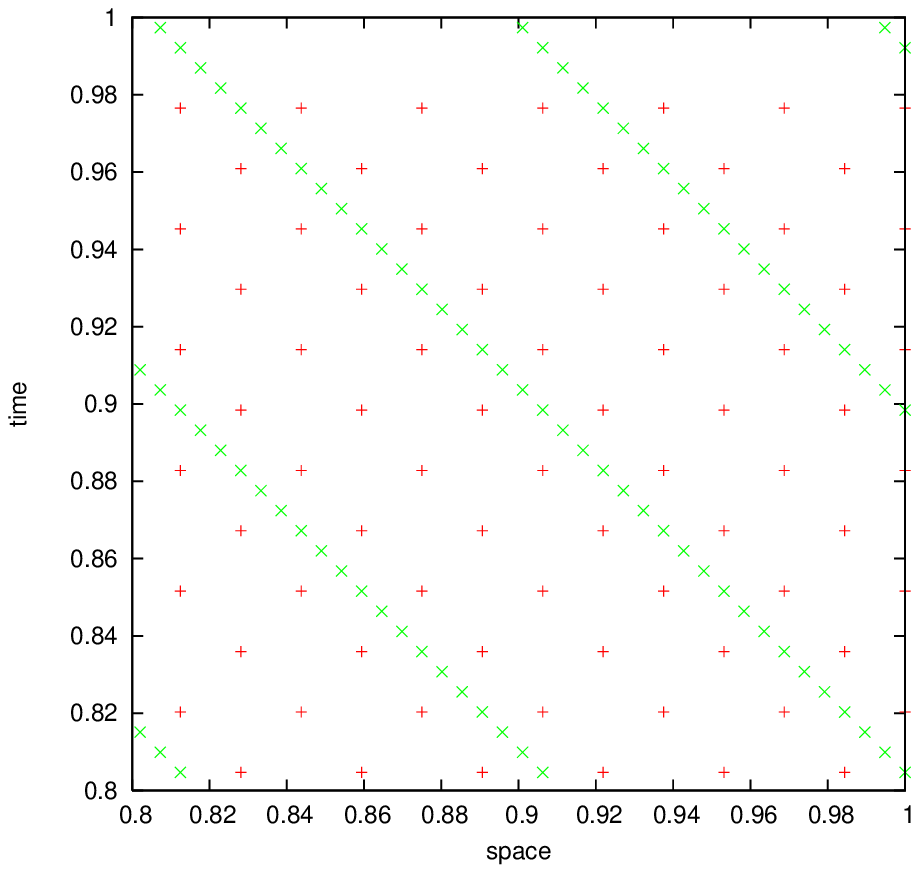}}
\end{center}
\caption{Two embeddings of a causal set into a spacetime.  (a) The causal set
  has 512 elements, and is faithfully embedded into a flat $2d$
  spacetime with cylindrical topology $S^1 \times I$.  The links are shown in
  green.
  (b) The figure depicts a (non-faithful) embedding into a region of $2d$
  Minkowski space.  The red and green distributions of points are related to
  each other by a boost of 4/5.}
\label{embedding}
\end{figure}

Figure \ref{embedding} illustrates two embeddings of causal sets into
spacetimes, one faithful and one not.  Some important features to note in
these diagrams regard the behavior of the \emph{links} in the causal sets, which
are those relations which are `irreducible', in that they are not implied by
transitivity.  For the faithful embedding on the left, one can note that (a)
the number of links 
connected to each element is very large, and (b) 
the vast majority of links connect elements from very distant regions of the
spacetime, in any given frame (such as that of the figure).  This is because
any given link only appears short in one frame, it will be long in all
others.  Since almost all frames are `another frame', almost all
links appear to be very long. 
In the embedding on the right, on the other hand (in which the links are not
shown explicitly), each element is connected to 
only
four links.  In the red frame the links connect nearby elements, while in the
green frame the ends of the link are becoming more distant in one direction
(and extremely close in the other).
This embedding fails the `faithfulness' condition, that the expected number of
elements mapped to any region is equal to its volume.  There are obviously
very large regularly shaped regions which fail this criterion.  In this way
regular lattices fail to be manifoldlike, if one does not go all the way to
the continuum ($\rho \to \infty$) limit.
Thus in this way the Lorentz invariance appears to in fact \emph{emerge} from
a discrete relational theory, as the only way in which the correspondence
between number of discrete elements and spacetime volume can be independent
of the region one is considering.

\subsection{Hauptvermutung} 

\begin{figure}[htbp]
\begin{center}
\psfrag{C}{$C$}
\psfrag{p1}{$\phi_1$}
\psfrag{p2}{$\phi_2$}
\psfrag{(M1,g1)}{$(M_1, g_1)$}
\psfrag{(M2,g2)}{$(M_2, g_2)$}
\scalebox{.78}{\includegraphics{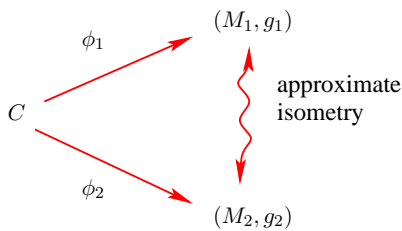}}
\cmb{8}{\caption{\label{hauptvermutung} `Hauptvermutung' of causal sets}}
\end{center}
\end{figure}

Named after the famous conjecture in topology, that every triangulable space
has a unique triangulation (which was later proved to be false), 
it is
conjectured that a causal set which faithfully embeds into a spacetime
manifold determines the manifold up to `approximate isomorphism'.  Somewhat
more precisely, we can define the `Hauptvermutung' (central conjecture) for causal sets as
follows (see figure \ref{hauptvermutung}).  
Given a causal set $C$, and two faithful embeddings $\phi_1 : C \to (M_1,
g_1)$ and $\phi_2 : C \to (M_2, g_2)$ from $C$ into spacetime manifolds, then
$(M_1, g_1)$ and $(M_2, g_2)$ must be approximately isometric.  The
`approximately' is necessary because of the discreteness --- the embedding is
obviously blind to continuum structures smaller than the discreteness scale.
The precise mathematical statement of the Hauptvermutung remains open,
however there does exist a body of work on defining a distance measure
between Lorentzian geometries \cite{noldus}.

It may seem a lot to hope, to recover the full spacetime geometry from the
discrete order.  However, the expectation is not unreasonable.  The
microscopic order more-or-less directly encodes the macroscopic causal
order.  By a theorem of David Malament and others
\cite{malament,hawkingkingmccarthy}, we know that the causal ordering of events
in 
spacetime contains enough information to
recover the topology, differential structure, and conformal metric.  Thus all that remains is the conformal factor,
which encodes volume information.  Here the discreteness plays a crucial
role, by providing the missing volume information, via the correspondence
between number and spacetime volume expressed in the faithfulness condition
on the embedding.  In the continuum one needs to add the gravitational field
to get geometry, while in the discrete the geometry arises naturally, without
needing to add any additional mathematical structures.
Thus it is not unreasonable to expect that one can recover the complete
spacetime geometry from the discrete causal order.
Note here that the Lorentzian signature arises naturally, as the only one
capable of distinguishing past from future.

\subsubsection{Dimension}
An obvious question which arises when discussing causal sets is how can one
predict the dimension of an approximating spacetime, given only the causal
set.  There have been a number of proposals for how to do this, for example
the midpoint scaling and Myrheim-Meyer dimension, which are both described in
\cite{valdivia,reid_dimension}.
The Myrheim-Meyer dimension \cite{myrheim,meyer_thesis} works by counting the number of relations in an order
interval, and comparing it with what one would get from sprinkling into
$\M^d$. 

In section \ref{manifoldlikeness} we sketch an alternate method for
estimating the dimension of causal sets, based upon the counting of $n$-links
(defined in section \ref{nlinks-sec}), and present some preliminary
results.

\subsubsection{Timelike distance}

\begin{figure}[htb]
  \begin{center}
    \scalebox{.8}{\includegraphics{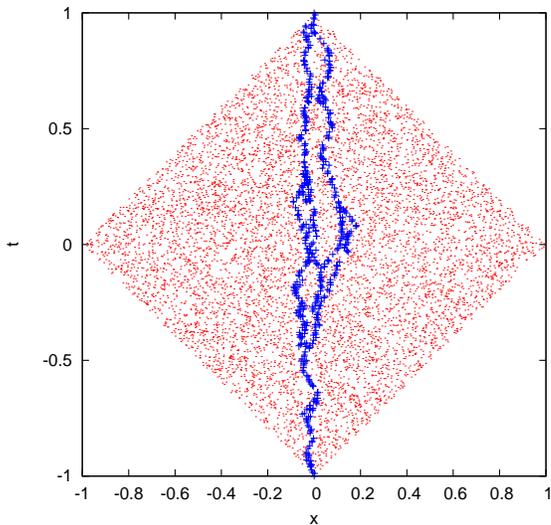}}
    \cmb{5.5}{\caption{\label{geodesics} The collection of longest chains
        (\emph{geodesics}) between a pair of elements in $\M^2$.}}
  \end{center}
\end{figure}

Since the fundamental relation defining the causal set is intrinsically
timelike, one can imagine that it is relatively straightforward to extract
timelike distances from the causal set.  The natural choice which is likely
extendable to curved spacetime is to count the number of links in the longest
chain connecting the two related elements $x \prec y$
\cite{myrheim,bachmat,joe_review}. (In general such a longest chain will be
far from unique.  Figure \ref{geodesics} shows the collection of longest
chains between a pair of elements in $\M^2$.)

By a theorem stated in \cite{bg}, we know that this length $L$
converges to the proper time separation between $x$ and $y$, in particular
\be
\frac{L}{(\rho V)^{1/d}} \to m_d
\ee
in probability as $\rho V \to \infty$.  Here $\rho$ is the sprinkling
density, $V$ is the volume of the interval $[x,y]=J^+(x) \cap J^-(y)$, $d$ is
the spacetime dimension, and $m_d$ is a constant depending only on the
dimension.  It's exact value is known only in two dimensions, $m_2=2$.
It's value in three dimensions was measured in \cite{diameters} to be $m_3 =
2.296 \pm .012$.
If we write the volume of the interval as $V = D_d T^d$, with $T$ the proper
time we seek, and $D_d$ a dimension dependent coefficient (we will need $D_3 = \frac{\pi}{12}$), then in the
infinite $\rho$ limit
\bne
T = \frac{L}{m_d \rho^{1/d} D_d^{1/d}} \;.
\label{T}
\ene

\begin{figure}[htbp]
  \begin{center}
  \psfrag{m3eff}{$\meff{3}$}
  \psfrag{log2 N}{$\log_2 \langle N \rangle$}
  \includegraphics{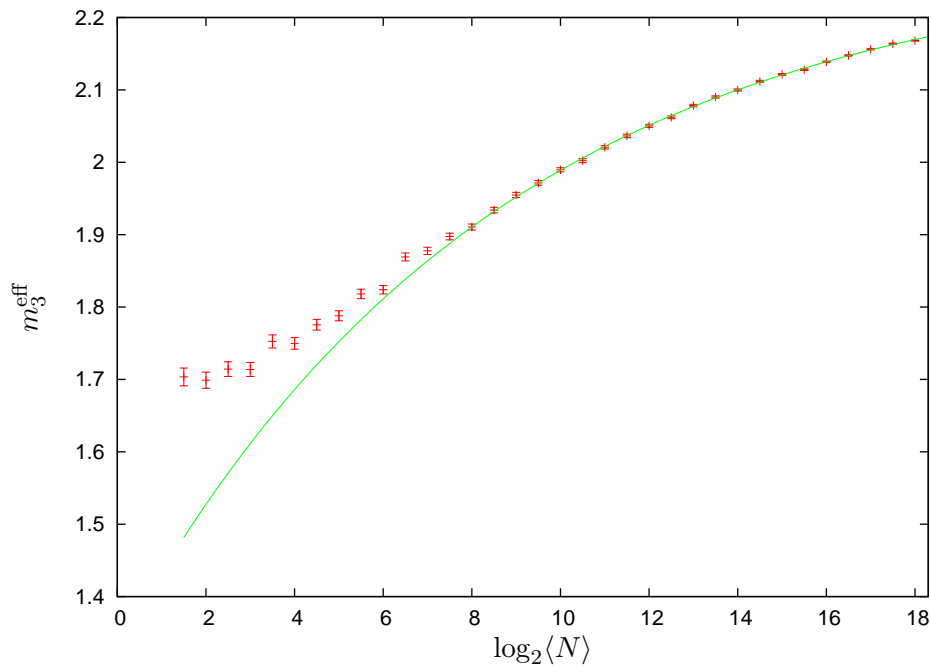}
  \end{center}
    \caption{\label{m3eff_fig} Convergence of length of longest chain $L$ to
      proper time $T$ for a faithful embedding into an interval of $\M^3$.
      The green fit to the asymptotic form measures the constant of
      proportionality $m_3$.}
\end{figure}

To get some feel for how well this works `in practice', we check it on a
computer, using code for the Cactus high performance computing framework
\cite{cactus} mentioned in \cite{diameters}.  We sprinkle into an
interval of height $T=2$ (in arbitrary units), count the length of the longest chain within that interval, and
from it compute, from (\ref{T}), an `effective' $m_d$
\bne
\meff{d} = \frac{L}{T(\rho D_d)^{1/d}}
\label{m3eff_eqn}
\ene
which is relevant for finite $\rho V$.  For $d=3$ we get the results depicted
in figure \ref{m3eff_fig}.
Each datapoint results from 1600 sprinklings into an interval of $\M^3$, each
with
a mean number of elements $\langle N \rangle$ (the mean of the Poisson
distribution) as indicated.  We plot the mean $\meff{3}$ from
(\ref{m3eff_eqn}), along with the standard estimate of its error.  The fit is
to the function $m_3 + a e^{b \log_2 \langle N \rangle}$, for the datapoints
$\langle N \rangle \geq
2^8$.  We get $m_3 = 2.2856 \pm 0.0063$, which is consistent with the
measurement in \cite{diameters} (that employs a slightly different
procedure)\footnote{Here we are effectively conditioning on the existence of
  sprinkled elements at the endpoints of the interval.  Since the occupation
  probability at each point in a Poisson process is independent of those at
  every other point, this should make no difference.  In \cite{diameters} the
  authors select a pair of elements that land nearest the endpoints of the
  interval, and use the interval formed by those two elements.  It appears
  that this distinction does not affect our measurement of $m_3$.}.

Note that for any finite region ($\rho V$ finite), the $\meff{3}$ is less
than the asymptotic value $m_3$, and thus a measurement of proper time using
the length of the longest chain via (\ref{T}) will always underestimate the
continuum proper time.  We call this effect \emph{timelike underestimation}.

\section{Early Prescription for Spacelike Distance}
\label{bgdist-sec}

We would like to recover the full spacetime geometry from the discrete causal
ordering.  We have seen above how to get dimension and timelike distances.
How can one extract spatial information?  How does spatial geometry come out?
We start with the simple question of spatial 
distance in flat spacetime.  Can
we recognize 
the spatial distance between elements of a causal set which can be faithfully embedded into Minkowski space?

An obvious first guess is to use the spacelike points to locate a timelike
pair, whose distance is the same as the spacelike distance, and then use the
timelike method above.  In the continuum, this can be done, as follows.
Consider a pair of spacelike separated elements $x$ and $y$, whose spatial
distance we would like to measure.  Consider
those pairs of points $w$ in the common past $J^-(x) \intersect J^-(y)$ and
$z$ in the common future $J^+(x) \intersect J^+(y)$ which minimize the
timelike distance (length of the longest chain $L_{wz}$) between $w$ and $z$.  In the continuum, every such pair is
separated by exactly the spacelike distance between $x$ and $y$. 

In the discrete causal set context, the above construction works fine for
causal sets which faithfully embed into $\M^d$, for $d<3$.  This is because
for $d=2$ there is a unique choice for the pair $(w, z)$ in the continuum
($w$ is the point at the intersection of the past light cones of $x$ and $y$,
and $z$ is the intersection of their future light cones).  For a causal set
sprinkled into $\M^2$, the pairs $(w, z)$ at minimum timelike separation
(length of the longest chain) will lie close to the corresponding `minimizing
pair' in the continuum.

At larger dimension $d$, in the continuum, one find a $d-2$ dimensional
submanifold of points $(w, z)$ at minimum timelike separation.  One way to
see this is that the pair $(x,y)$ breaks the Lorentz symmetry in one
direction, but the picture is still invariant under boosts in any orthogonal
direction.  Starting from any arbitrary frame, in which the pair $(w, z)$
both occur at the centroid of $x$ and $y$, one can trace out the submanifold
by applying all 
boosts which leave $x$ and $y$ invariant.

\begin{figure}[htbp]
\begin{center}
\cmt{4.4}{
\includegraphics[width=4.3cm]{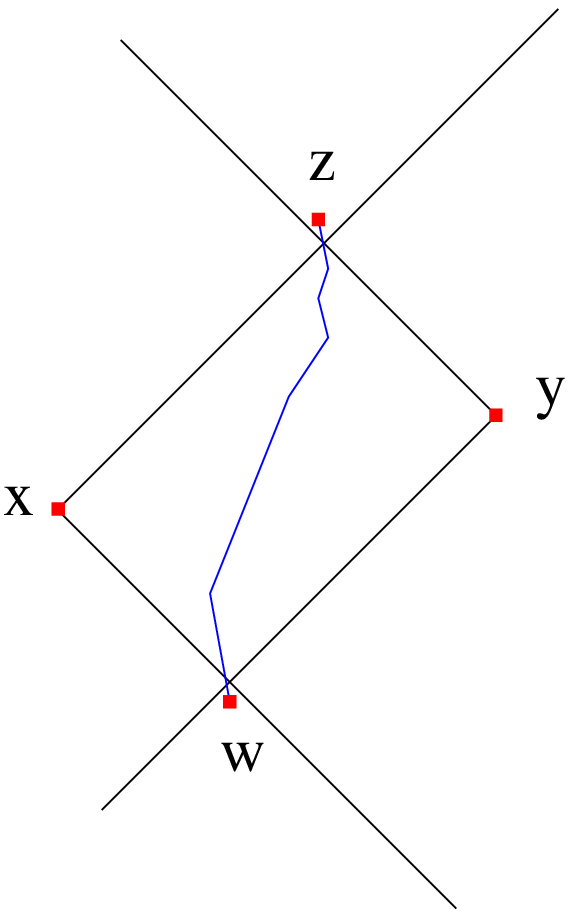}
}
\cmt{9}{
  \psfrag{w1}{$w_1$}
  \psfrag{w2}{$w_2$}
  \psfrag{w3}{$w_3$}
  \psfrag{z1}{$z_1$}
  \psfrag{z2}{$z_2$}
  \psfrag{z3}{$z_3$}
  ~\vspace{-65mm}\\ \includegraphics[width=11.5cm]{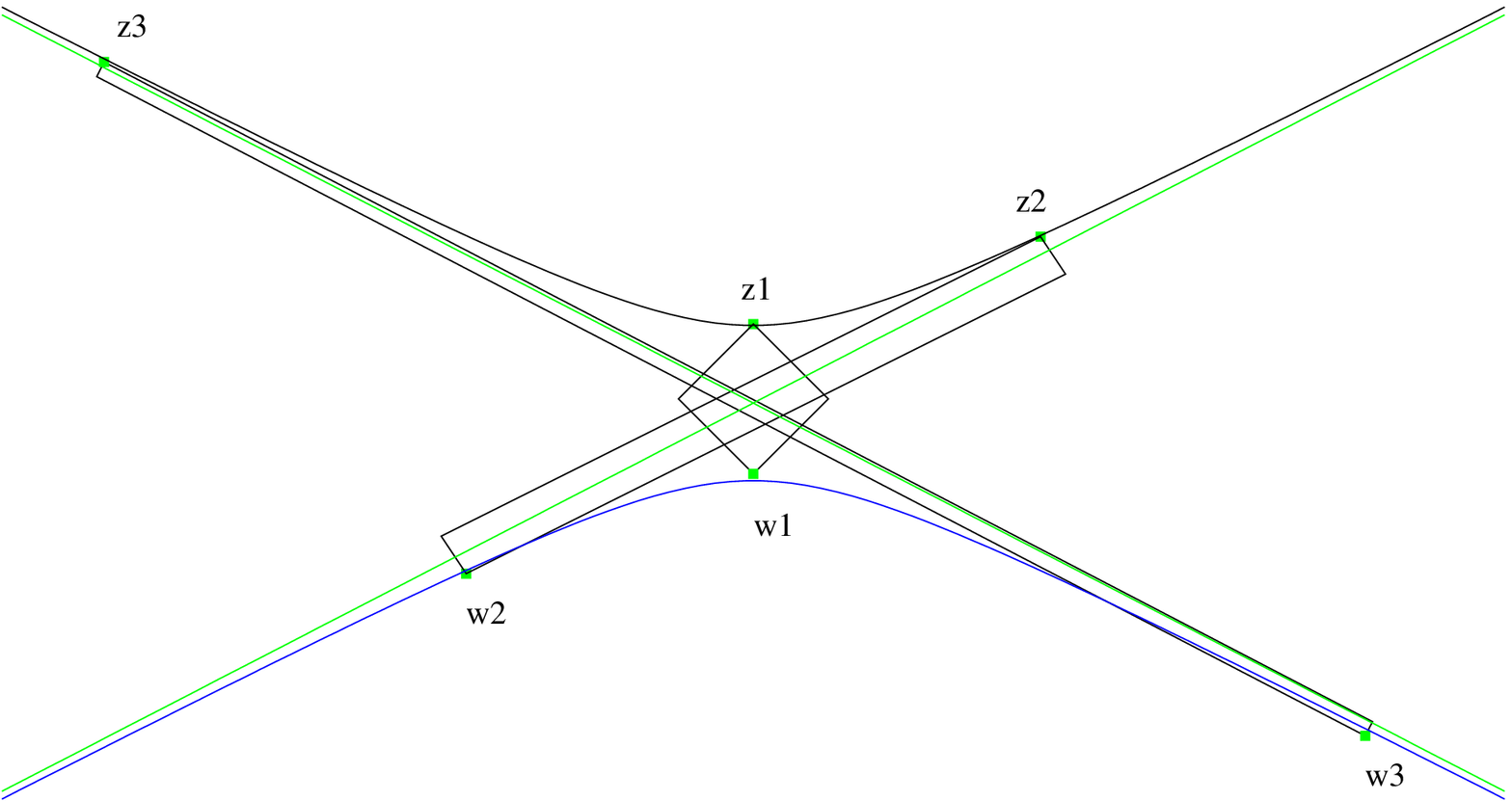} }
\end{center}
\hspace{5mm} (a) \hspace{2.3in} (b)
\caption{(a) Two spacelike elements $x, y$ in $\M^3$, and a pair $w,z$ in
  their common past/future.
  (b) The elements $x$ and $y$ lie at the center of the figure, but
    displaced out of and into the page by some finite equal amount.  The
    hyperbolae indicate the intersection of the past and future light cones of
    $x$ and $y$.  The long straight green lines are the asymptotes of the
    hyperbolae.  The green dots with black intervals indicate the projection
    of the causal intervals for minimizing pairs onto the plane of the page.}
\label{bgfailure_fig}
\end{figure}

Unfortunately this infinity of `minimizing pairs' causes problems in the
discrete case of a random sprinkling, because it causes one to consider an
infinite number of independent regions when computing the spatial distance
between $x$ and $y$.  To understand why this simple prescription for
spacelike distance fails in dimensions greater than 1+1, consider the
spacelike pair $(x, y)$ in figure \ref{bgfailure_fig}(a).  Shown are a pair
$(w,z)$ (a \emph{minimizing pair}) which are close to the intersection of $x$ and $y$'s future and past
light cones, and might locally minimize the length of the longest chain
between $w$ and $z$.  However, our distance prescription seeks a global
minimum, not a local minimum, so we must consider `what happens in other
frames' as well.  Figure \ref{bgfailure_fig}(b) depicts the same situation, in the
frame in which $x$ and $y$ are simultaneous, and $w$ and $z$ occur at the
same spatial location.  ($x$ and $y$ are not shown explicitly in the diagram;
they are displaced into and out of the page by a small amount.)  
Shown are three candidates for pairs $(w,z)$, along with the order interval
between them, which is the portion of the causal set which determines the
length of the longest chain between them.
If we compare the intervals for two pairs which are `highly boosted with
respect to each other' (meaning the boost parameter relating the frames in
which the pairs $(w,z)$ are simultaneous) have very little overlap.  But each
of these intervals contains the same spacetime volume of Minkowski space.
The sprinklings in each of these regions is independent (save the tiny region
of overlap), since the Poisson distribution is independent at each point.
Since there are an infinite number of candidate pairs $(w,z)$, to find the
global minimum, we must sample the Poisson distribution an infinite number of
times, to get the portion of the causal set in each of these independent
regions.  One possible result for this sample, which has finite probability,
is that the region is empty.  
Thus this event (of finding a candidate pair $(w,z)$ whose interval $[w,z]$
is empty of sprinkled elements) must occur an infinite number of times, and
so the distance between $x$ and $y$ will always be exactly two\footnote{two
  because by definition $w \prec x \prec z$ and $w \prec y \prec z$ (and $x
  \nprec y$), each of which are chains of length 2.}, regardless of
where they fall in Minkowski space.

\subsection{Numerical evidence} 
\label{bgfailure-sec}
The above argument depends upon a sprinkling into the entirety of Minkowski
space.  What if one has a sprinkling into a finite region of Minkowski, such
as might be physically realistic in a finite universe, or as can be simulated
on a computer?  Is it possible to see the above degeneracy in spatial
distance, while considering only that portion of the causal set which is
contained in a finite portion of Minkowski space?  The probability to see a
region of spacetime volume $V$ which is empty of sprinkled elements is
$e^{-V}$.  In $\M^3$, 
an interval of height $T=4$ has volume $V = \frac{\pi}{12}4^3 \approx 16.755$,
and
the probability to find an empty interval of height $T=4$ is $5 \times 10^{-8}$.
Thus a lower bound on the size of a region of $\M^3$ which would be required
to see this degenerate distance is $17/5 \times 10^{-8} \approx 3 \times
10^8$.

\begin{figure}[htb]
\begin{center}
\psfrag{8}{$\Delta$}
\includegraphics[width=7.7cm]{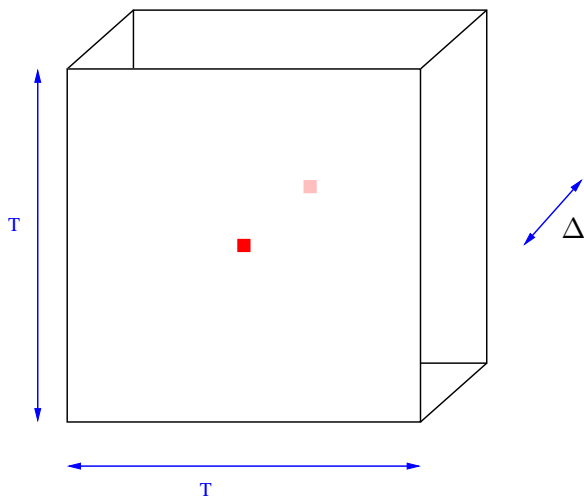}
\begin{minipage}[b]{18pc}
\caption{Growing region of $\M^3$ into which we sprinkle.  $\Delta$ is held
  fixed to a small value (4 or 8), while $T$ is increased as far as the
  computer allows ($2^{7.75} \approx 215$).}
\end{minipage}
\end{center}
\label{sprinkling_box}
\end{figure}
To explore this on the computer, we sprinkle into a $T \times T \times
\Delta$ box, as illustrated in figure \ref{sprinkling_box}.  In this section we
use units such that $\rho \equiv 1$.  The spacelike pair
$(x,y)$ are placed at the small red squares, which lie in the center of the two
largest faces as shown.
  In the simulations, we hold the
distance $\Delta$ between $x$ and $y$ fixed, while increasing the size of the
box in the lateral directions.  If the distance measure works well, then
obviously increasing the size of the box in the lateral directions should
have no affect.
\begin{figure}[htb]
\begin{center}
  \psfrag{log2 T}{$\log_2 T$}
  \psfrag{deviation}{$\left<\frac{X - \Delta}{\Delta}\right>$}
  \psfrag{8}{\hspace{-9mm}$\Delta=8$}
  \psfrag{4}{\hspace{-9mm}$\Delta=4$}
  \includegraphics{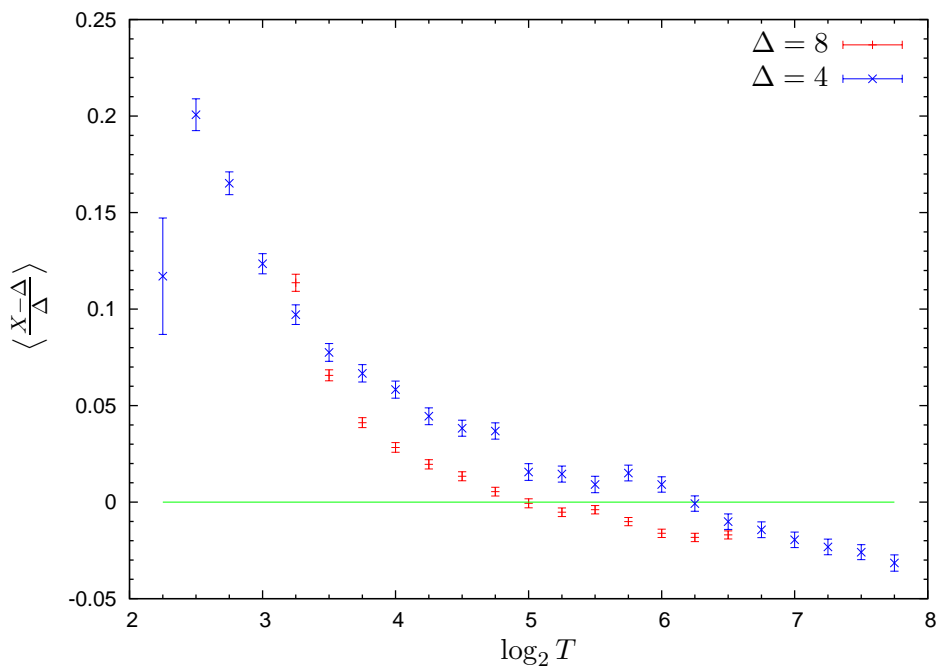}
\begin{minipage}[b]{7.5pc}
\caption{Demonstration of the degeneracy of the `naive' prescription for
  spacelike distance of section \ref{bgdist-sec}.  Each data point gives the
  mean and standard error for 900 sprinklings.}
\end{minipage}
\end{center}
\label{bgfailure-fig}
\end{figure}
We see in figure \ref{bgfailure-fig} that this is not the case.
There we show the results from two simulations, one with $\Delta=4$, and one
for a more distant pair $\Delta=8$.  $X$ is the smallest value of (\ref{T})
for a pair of elements $(w,z)$ in the common past and future of $x$ and $y$.
In all these simulations we use a $\meff{3} = 1.75815$, which we carefully
measure to be the appropriate value for an interval of size $\rho
V=\frac{\pi}{12}4^3$.  This eliminates the effect of timelike underestimation.

At $T=\Delta$ the common future and
past of $(x,y)$ are outside of the sprinkling region, so the spatial distance is
undefined.  For slightly larger $T$, we observe a substantial overestimation
of the continuum distance $\Delta$.  This occurs because it is unlikely that
a sprinkled element will land close to the intersection of $x$ and $y$'s
light cones.  The pair $(w,z)$ will always be separated by a continuum
(timelike) distance $> \Delta$, hence the overestimation.
This overestimation will always be of the order of the discreteness scale,
and so, even though in this context is rather large, will be irrelevant for
macroscopic distances.

At large $T$, we see the overestimation decrease, past zero to negative
values.  Since the use of $\meff{3}$ already corrects for the timelike
underestimation effect, this decrease must be due to the degeneracy of
the spacelike distance definition of section \ref{bgdist-sec}.  If we could
expand $T$ all the way to infinity, we expect to see the deviation decrease
to $\frac{\frac{2}{D_3^{1/3} \meff{3}} - \Delta}{\Delta} \approx -.55544$ for
$\Delta=4$.

~\\
It appears that this naive prescription for spatial distance on a causal set
fails, due to the Lorentz invariant nature of the embedding, and the
random nature of the Poisson process.  How might one deduce spatial distances
on a causal set?  One possibility is to select enough additional elements to
eliminate the boost freedom mentioned above.  
This leads to more of a `diameter' measure, than a distance measure, in that
it can be interpreted as giving the diameters of spheres defined by the $d$
elements.  This is described in section \ref{diameters-sec}.

An alternate proposal is to take advantage of the fact that links in a causal
set closely track the light cones, and this can be used to locate appropriate
pairs $(w,z)$ without resorting to a global minimum.  This leads to the
2-link distance proposal of section \ref{2links}.

\section{Diameter measures}
\label{diameters-sec}

In order to locate a relatively unique pair $(w,z)$, one can select, instead
of simply a pair of elements $(x,y)$, a collection of $d$ mutually unrelated
elements (an \emph{antichain}) in $\M^d$.  One can then select a $w$ which
is in the common past of the entire antichain, and a $z$ which is in the
common future, and seek a pair which minimizes the length of the longest
chain between $w$ and $z$.  It turns out that this measures the diameter of
the smallest $d-1$-ball, which lives on the surface of simultaneity of all
$d$ elements of the antichain, and contains them all.  It is called $l_g$ in
\cite{diameters}.

Using the same antichain, one can also construct a $d-2$-sphere, which lives
on the same spatial hypersurface, and contains each of the $d$ elements of
the antichain.  It is possible to write down an expression in terms of order
invariants which measures the diameter of this sphere.  It is called $l_s$ in
\cite{diameters}.  Both of these diameter measures are described there in
detail, along with a number of possible applications.

\section{n-links}
\label{nlinks-sec}
\begin{figure}[htb]
\begin{center}
\includegraphics{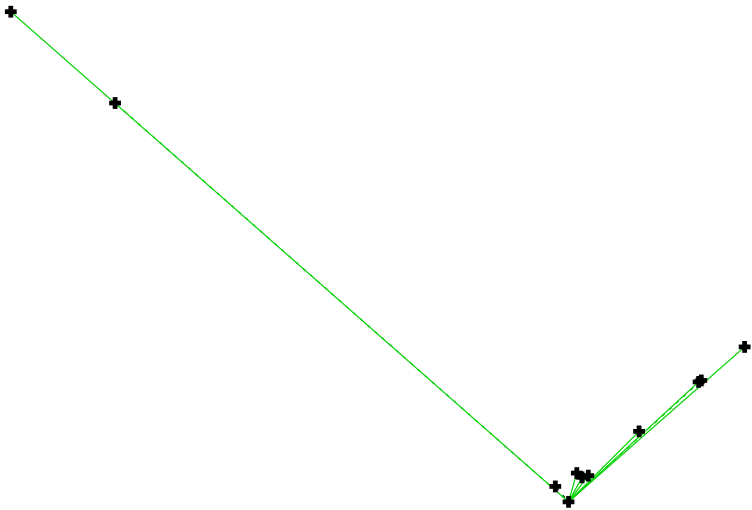}
\cmb{5}{\caption{\label{links-fig}
Links to the future of an element in Minkowski space.}}
\end{center}
\end{figure}

As mentioned above, an alternate approach to measuring spatial distances in a
causal set is to use the fact that the \emph{links} of the causal set lie
very close to the light cone in a faithful embedding, and effectively give
the light cone structure of the causet.  One way to see this is that the
proper time between a pair of related elements is measured by the length of
the longest chain connecting them.  A null relation corresponds to a minimum
proper time (zero), so the analogue in a causal set should correspond to a
minimum length of the longest chain, which is a single 
link.  Figure \ref{links-fig} depicts some of the links to the future of an
element in Minkowski space.

\begin{figure}[htbp]
\begin{center}
\includegraphics[width=8cm]{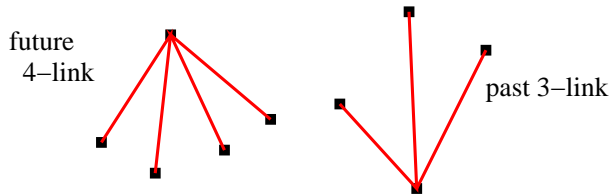}
\cmb{6}{\caption{\label{nlinks-fig} Illustration of $n$-links.}}
\end{center}
\end{figure}

We define an \emph{$n$-link} as an element which is linked to each member of an
$n$ element antichain. 
If this
element is to the future (past) it is called a future (past) $n$-link.
Some examples are depicted in figure \ref{nlinks-fig}.
Given the above discussion, it is fairly clear that an $n$-link corresponds
to the intersection of light cones emanating from each element of the antichain.

How many $n$-links does one expect to find, to the future of a given
$n$-antichain (an antichain with $n$ elements), for a causal set faithfully embedded into $\M^d$?  For $n<d$,
one expects an infinite number, for similar reasons as discussed in section
\ref{bgdist-sec}.  Roughly speaking $n$ elements are not enough to select a surface of
simultaneity, so in each of an infinite number of frames one expects to see
an $n$-link with some finite probability.  For $n>d$ there should not be any
$n$-links, unless the $n$ elements are arranged in some special way (e.g.\ as
in figure \ref{links-fig}, which depicts a 9-link within a sprinkling of $\M^2$).  For
$n=d$, the intersection of the future light cones consists of a single
point.  The probability that there is an element near (and to the future of)
this point is moderate, but the probability that it is linked to each element
of the antichain diminishes rapidly with their distance.

One application of $n$-links in causal sets is they can provide a definition
of spatial distance which is devoid of the degeneracy from which the
definition given in section \ref{bgdist-sec} suffers.  This will be explained
in detail in the next section.

Another application may be that they can be used as in indicator of
`manifoldlikeness' in a causal set (cf. \cite{homnum}).  The behavior of the
number of $n$-links in a causal set that is well approximated by Minkowski
space is sketched above, which is potentially quite specific given that it
should hold for all $n$.  For example, if one does have a causal set which
faithfully embeds into $\M^d$, then the above behavior should fairly clearly
pick out the value of $d$.  This could compared with other dimension
estimators as a further test.  An initial attempt at identifying
manifoldlikeness in this way appears in section \ref{manifoldlikeness} below.

\section{2-link Distance}
\label{2links}
The definition of spacelike distance given in section \ref{bgdist-sec} fails
because for any spacelike pair $(x,y)$ there are an infinite number of
potential `minimizing pairs' $(w,z)$ which lie close to the intersections of
the light cones of $x$ and $y$.  Since one takes a global minimum over all
such pairs $(w,z)$, one gets a trivial result.  If one could take an average
of `minimizing pairs' $(w,z)$, rather than minimum, this problem could be
avoided.  The problem is how to locate appropriate pairs of elements close to
the intersection of the light cones of $x$ and $y$, however this is exactly
what is provided by 2-links.

We thus define the 2-link distance between two unrelated elements $x$ and $y$
by the following algorithm:
\begin{enumerate}
\item Compute the set of all future 2-links of $x$ and $y$.
\item For each future 2-link $z$, compute the smallest distance $L_i$ from an
  element $w \in \past(x) \intersect \past(y)$ to $z$.
\item Store $L_i$.
\item Repeat the above with `future' and `past' interchanged.
\item Compute the average of the $L_i$.
\end{enumerate}

\begin{figure}[htb]
\begin{center}
  \psfrag{log2 T}{$\log_2 T$}
  \psfrag{deviation}{$\left<\frac{X - \Delta}{\Delta}\right>$}
  \psfrag{D = 8}{\hspace{-4mm}$\Delta=8$}
  \psfrag{D = 4}{\hspace{-4mm}$\Delta=4$}
  \psfrag{2-link distance}{\hspace{-7mm}2-link distance}
  \includegraphics{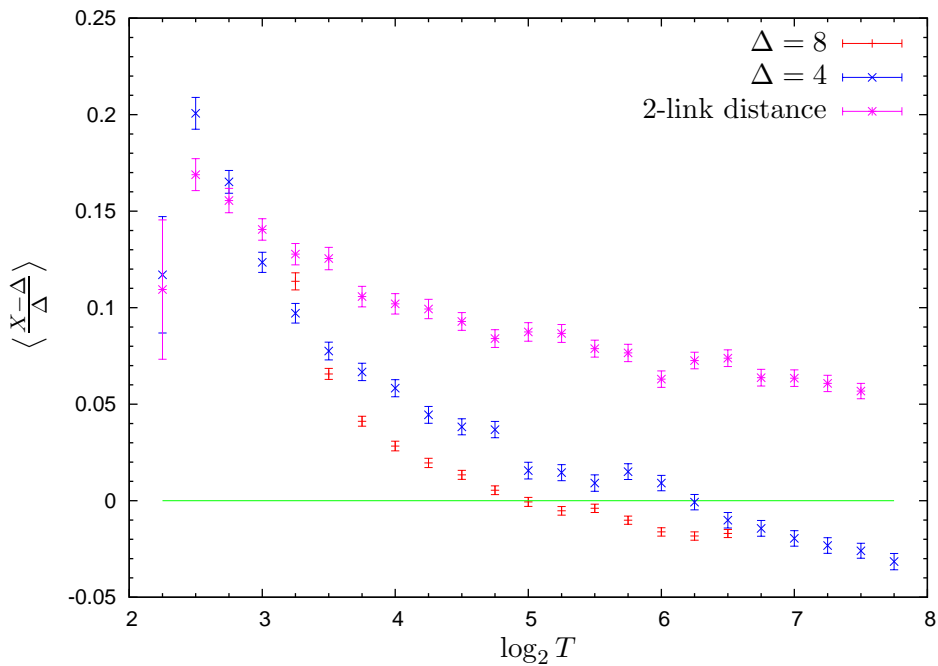}
\begin{minipage}[b]{7.55pc}
\caption{\label{2linkdist}
Comparison of 2-link distance at $\Delta=4$ (magenta) with the `naive' prescription for
  spacelike distance of section \ref{bgdist-sec} (red and blue).}
\end{minipage}
\end{center}
\end{figure}
In figure \ref{2linkdist} we contrast 2-link distance against the `naive'
spatial distance of section \ref{bgdist-sec}.  Each data point gives the mean
and its error, for 900 sprinklings into the boxes of figure
\ref{sprinkling_box}.  Again we see the large
spacelike overestimation for the 2-links, as for the naive spatial distance.
As the size of the box is increased, its accuracy increases slightly because
it finds more 2-links with which to estimate the distance.  The random
fluctuations in this figure are smaller than those of figure 14 of
\cite{diameters} because here we include past 2-links as well as future, and
also condition on there being 
causet elements at the center of the faces of the box. 

\begin{figure}[htb]
\psfrag{num 2-links}{mean number of 2-links}
\psfrag{box size}{box size $T$}
\begin{center}
\includegraphics{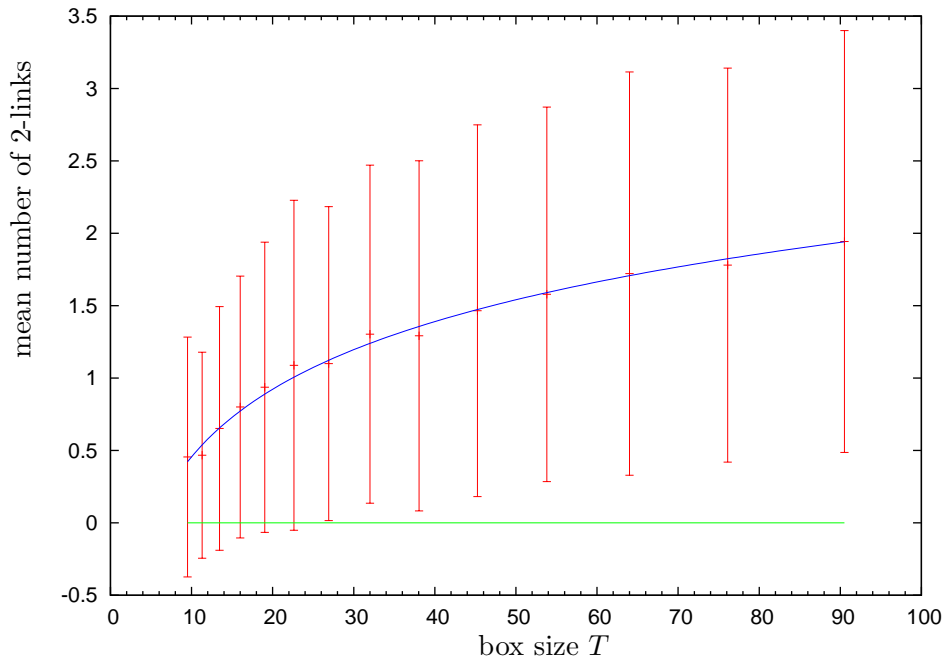}
\begin{minipage}[b]{7.55pc}
\caption{\label{num2links3d}
  Expected number of 2-links in the sequence of boxes of figure
  \ref{sprinkling_box}.  Each data point gives the mean and standard
  deviation from 900 sprinklings.}
\end{minipage}
\end{center}
\end{figure}

\begin{figure}[htb]
\psfrag{num 2-links}{mean number of 2-links}
\psfrag{box size}{box size $T$}
\begin{center}
\includegraphics{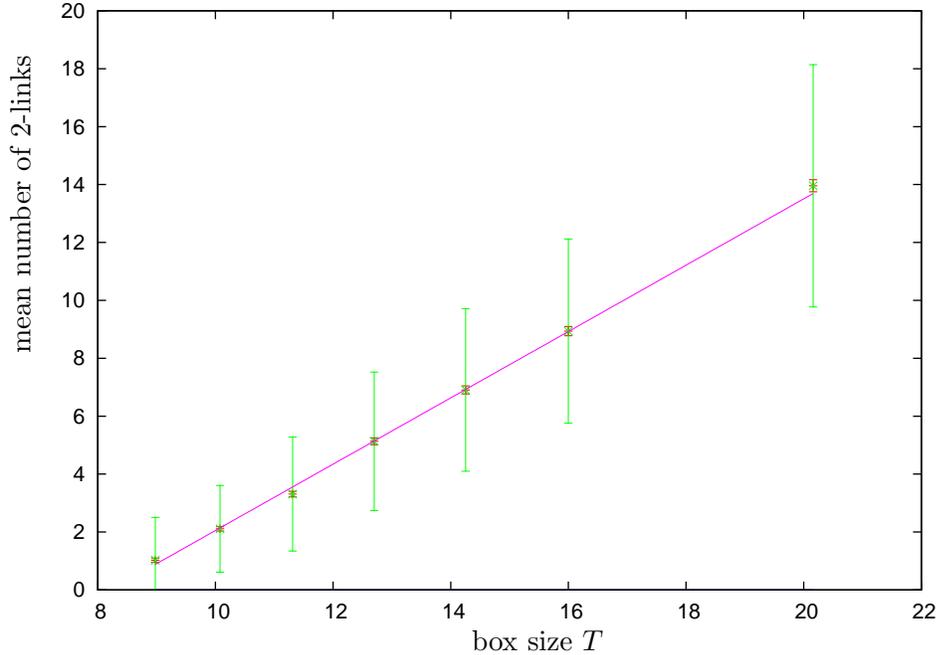}
\begin{minipage}[b]{7.55pc}
  \caption{\label{num2links4d} Expected number of 2-links in a sequence of
    boxes similar to figure \ref{sprinkling_box}, but with an added spatial
    dimension of size $T$.  Here each data point gives the mean, standard
    deviation, and standard error, from 403 sprinklings.}
\end{minipage}
\end{center}
\end{figure}

One may wonder how common are 2-links, if we are
to rely on their existence in order to define spatial distance in a finite region.  We
expect to find an infinite number in dimensions $> 2$, but this assumes
infinite Minkowski space.  Will there exist many in a reasonably sized
portion of our universe?  We address this question by searching for 2-links,
using the same computation as 
described in section \ref{bgfailure-sec}, with $\Delta = 8$.
In this case, however, we simply count the number of (future and past)
2-links, rather than compute the 2-link distance itself.  The results are
shown in figure \ref{num2links3d}.
The error bars indicate the usual estimate of the standard deviation.  Given
that they overlap with, or are not far from extending to, zero, there will be
a number of sprinkled causets which do not find any relevant 2-links.  In
these cases we simply discard the causet, though we do count it toward the
total number generated (900).
The
blue curve is a fit to $a + b \ln \langle T \rangle$.
In figure \ref{num2links4d} we count the number of 2-links in a rectangular $T
\times 8 \times T \times T$ box.  The spacelike elements $(x,y)$ are placed
in the centers of the $x$-faces, as before.  Here the number of 2-links grows
linearly with the box size, as one might expect.
The fitting function is $a + b \langle T \rangle$.  Here we show error bars
for both the
standard deviation and the error in the mean.

\section{Spatial nearest neighbors and curved spacetime}

One can use the 2-link distance in the context of curved spacetime, however
it has the unfortunate property that the distance between $x$ and $y$ depends
upon portions of the causal set which are arbitrarily `distant' from $x$ and
$y$.  To measure a length in my lab I must in principle consider the happenings in
Andromeda millions of years ago and also millions of years from now.
Given that we know that spacetime is locally flat, 
we can truncate the
search for 2-links at some intermediate mesoscale, beyond which we expect
curvature effects to become relevant.  It may be that the introduction of a
mesoscale is inevitable in discrete quantum gravity, as it arises in a number
of quite different contexts \cite{dalembertian,homnum}.

In order to get a reasonable distance measure from the 2-link distance, one
simply needs to find a `reasonable' number of 2-links.  One can thus define
an `infrared cutoff' in terms of some number of 2-links. 
This (subset of 2-links) of
course depends upon an arbitrary choice of frame, but this is not a problem,
in that the resulting 2-link distance should be independent of this choice,
provided that the region of the causal set that they `enclose' is well approximated by flat spacetime.

\begin{figure}[hbt]
\scalebox{2.10}{
  \psfrag{ 0.8}{}
  \psfrag{ 0.6}{}
  \psfrag{ 0.4}{}
  \psfrag{ 0.2}{}
  \psfrag{ 0}{}
  \psfrag{-0.2}{}
  \psfrag{-0.4}{}
  \psfrag{-0.6}{}
  \psfrag{-0.8}{}
  \includegraphics{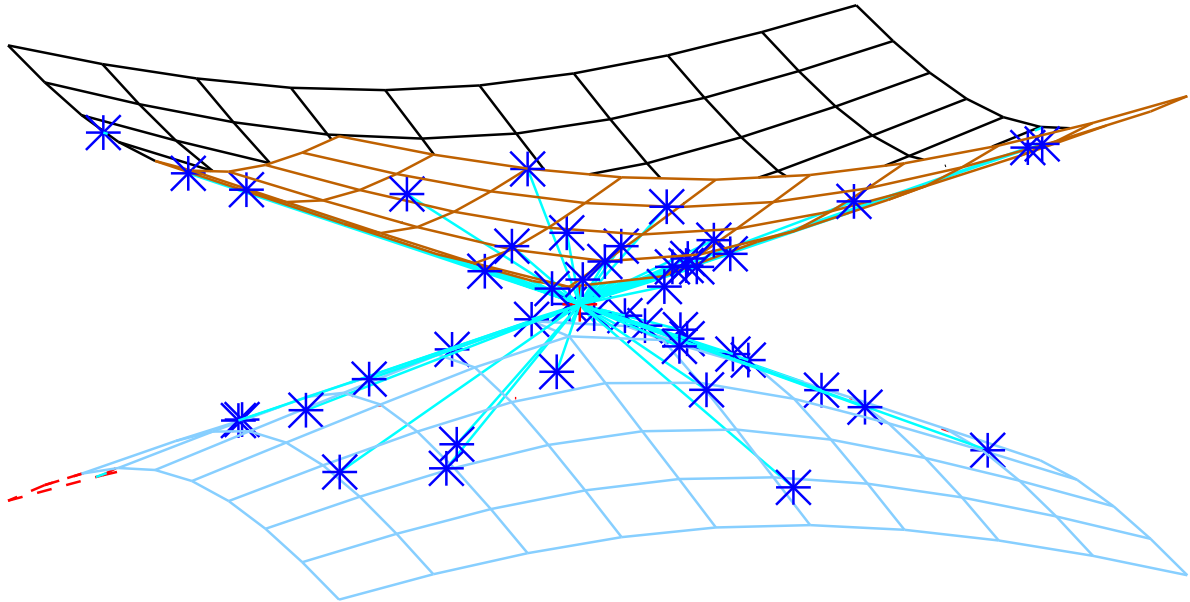}}
  \caption{Spatial nearest neighbors of an element in a sprinkling into (a
    fixed cube in) $\M^3$.  $\langle N \rangle = 65\,536$.  The future and
    past light cones of the `origin element' $x$ are shown.  The spacelike
    cyan lines are drawn between $x$ and each neighbor, for emphasis.}
  \label{neighbors}
\end{figure}

A proposal to address 
the need for the region enclosed by the `2-link minimizing pairs' to be flat
is to employ the 2-link distance to locate
`spatial nearest neighbors' in the causal set.  These would consist of
unrelated pairs of elements whose 2-link distance is beneath some threshold.
If the threshold is chosen small enough, then it can be easy to find enough
2-links with a quite modest cutoff.
Such nearest neighbors of an element within a sprinkling of
$\M^3$ is shown in figure \ref{neighbors}.

With such a spatial nearest neighbor relation in hand, one can define
distances of curves in a curved spacetime setting.  See \cite{diameters} for
some illustrations of spatial nearest neighbors in Minkowski space,
including an adjacency graph of a spatial slice, derived from a sprinkled
causal set.

\section{$n$-links as manifoldlikeness test}
\label{manifoldlikeness}

It is noted in \cite{diameters}, and also above in section \ref{nlinks-sec},
that the particular behavior of the number of $n$-links `attached'
to a given
antichain for sprinklings into Minkowski space may be used as an indicator
for `manifoldlikeness'.  By this we mean that, if the numbers of $n$-links
look like what one finds for Minkowski space, with many $n$-links for $n<d$,
few for $n=d$, and almost none for $n>d$, then the causal set may be likely
to faithfully embed into $\M^d$.

To get some idea how this might work in practice, we preform a
preliminary study, by counting 1- and 2-links
in a number of (finite) causal sets.  We find that the counts of 1- and
2-links are easily able to distinguish obviously non-manifoldike causal sets
from sprinklings, but they have difficulty distinguishing causal sets
generated by the sequential growth dynamics of \cite{cosacc}
from sprinklings into $\M^d$.  Physically, one might think of an abundance of
1-links as indicating the potential existence of light cones, as one finds in
spacetime, while an abundance of 2-links may indicate the existence of a
spatial direction orthogonal to the pair of elements in question.  3-links
could indicate spatial directions orthogonal to a plane, and so on.

\begin{figure}[h]
\begin{center}
\begin{minipage}{12pc}
\includegraphics[width=12pc]{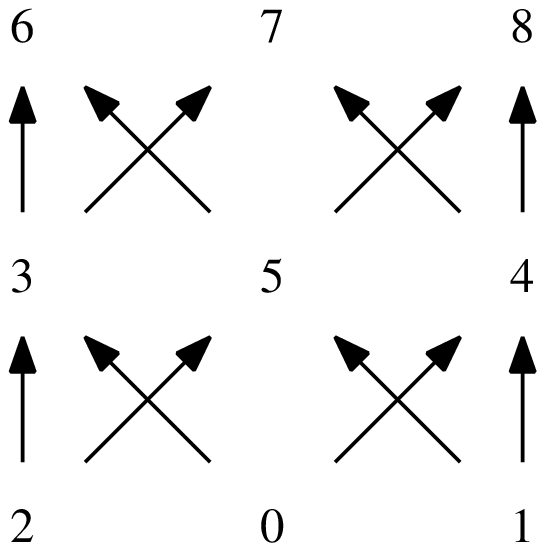}
\caption{\label{2_3crowns} A tower of three 3-crowns.  The numeric labels are
  arbitrary and have no significance.}
\end{minipage}\hspace{2pc}\begin{minipage}{10.0cm}
\includegraphics[width=9.1cm]{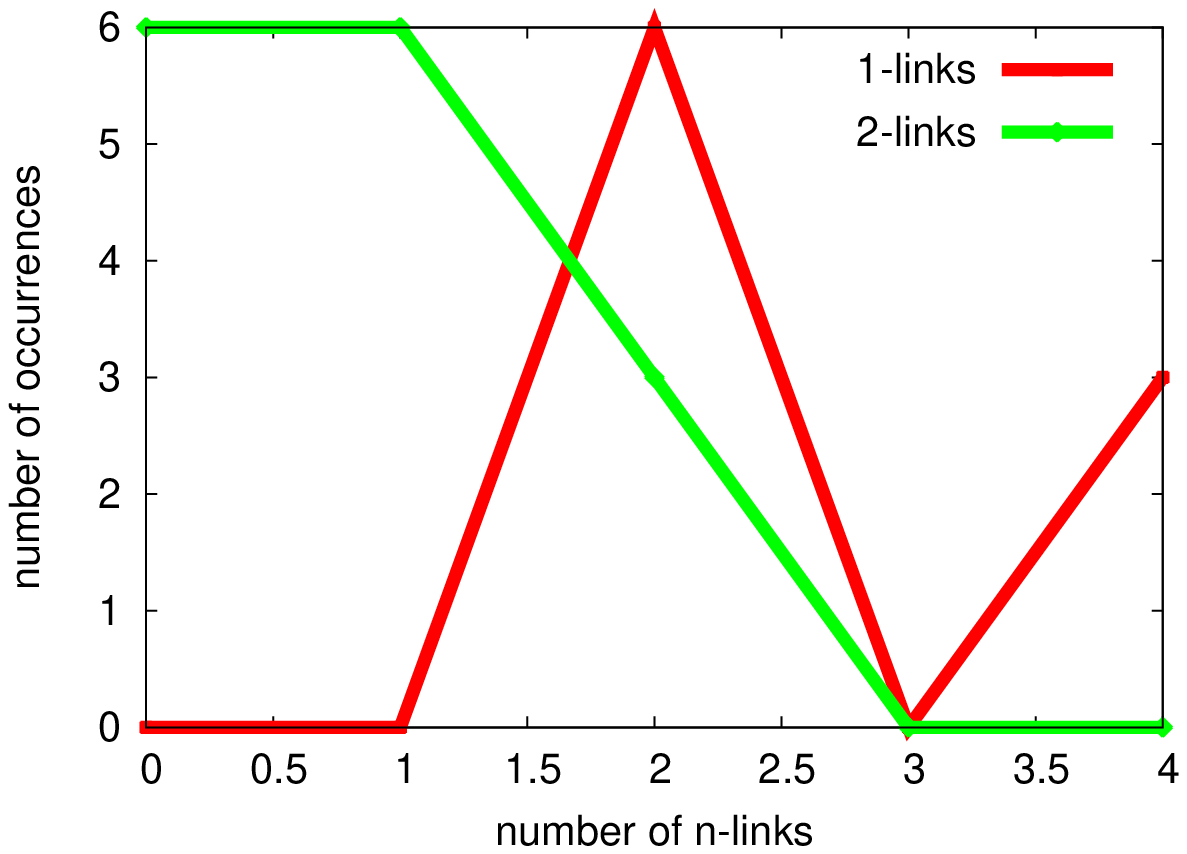}
\caption{\label{sample_hist}1- and 2-link histograms for the causal set in
  figure \ref{2_3crowns}.  Lines are drawn between the data points for clarity.}
\end{minipage}
\end{center} 
\end{figure}

The particular computation we perform is as follows.  Given a causal set $C$,
for each element $x$, we count the number of links $l$ connected to $x$, and
then form a histogram from these counts $l$.  We do likewise for 2-links: for
every 2-element antichain in $C$ we count the number of attached 2-links, and
form a histogram of these counts.
For example, for the
causal set in figure \ref{2_3crowns}, there are 
six elements with 2 (1-)links (the minimal and maximal elements (those with
an empty past and future respectively)), and three
with 4 links (those in the middle layer).  For 2-links, there are three pairs
with 2 2-links (those consisting of elements in the middle layer), six with
one 2-link (3 pairs each in the top and bottom layers), and six with zero
2-links (these come from pairs with elements in different layers).
The corresponding histogram is displayed in figure \ref{sample_hist}.

We begin with some sprinklings into intervals of Minkowski space, to see how
the counts of 1- and 2-links behave for manifoldlike causal sets.  
\begin{figure}[htbp]
\begin{center}
\psfrag{M2int}{$\M^2$}
\psfrag{M3int}{$\M^3$}
\psfrag{M4int}{$\M^4$}
\psfrag{M5int}{$\M^5$}
\psfrag{frw}{\hspace{-5mm}FRW}
\includegraphics[width=\textwidth]{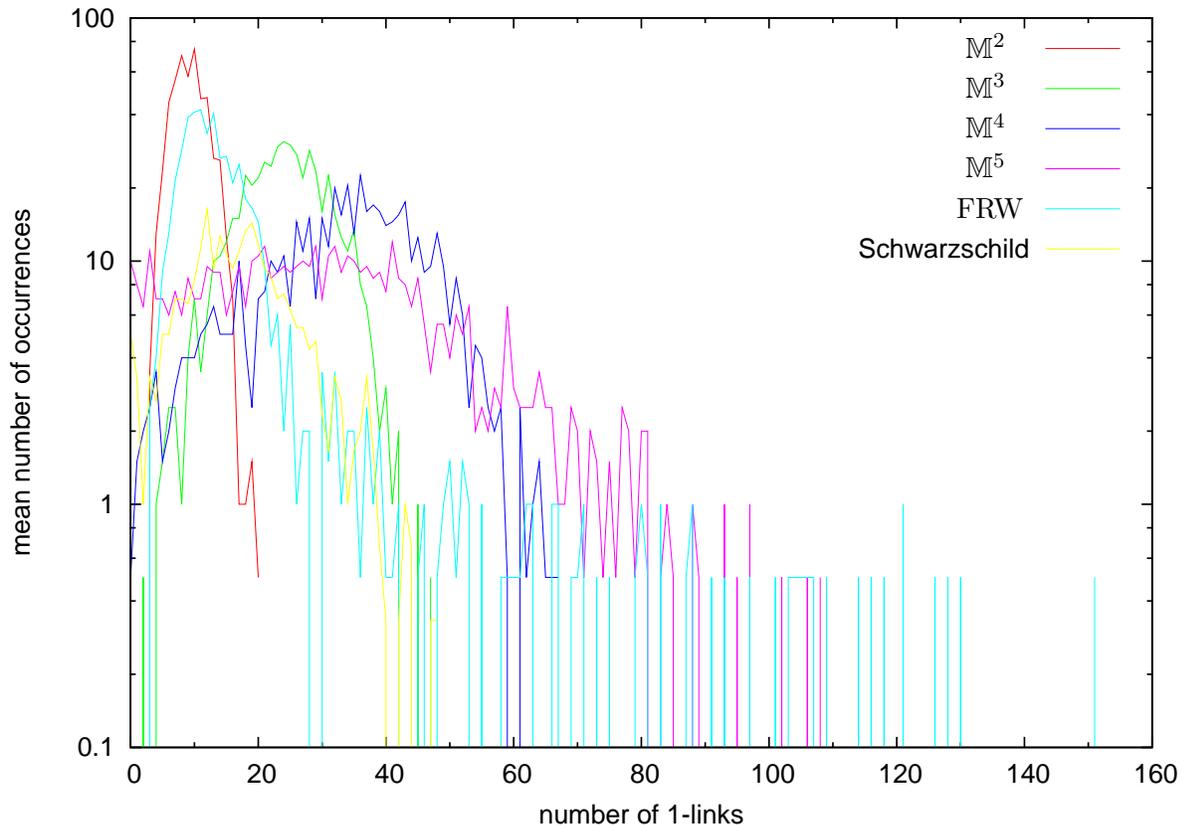}
\end{center}
\caption{Abundances of (1-)links in some manifoldlike causal sets.  Each
  histogram gives the mean from two sprinklings.}
\label{manifolds-1l}
\end{figure}
Figure \ref{manifolds-1l} shows histograms of the number of links attached to
a causet element, for sprinklings into six different spacetimes.
Four are
intervals in Minkowski space, of various dimensions.  The fifth is a
region of a conformally flat Friedman Robertson Walker universe, which
contains the initial singularity.  (The region is $\eta, x_1, x_2, x_3 \in
[0,1]$, in which coordinates the metric is $ds^2 = \eta^4(-d\eta^2 + dx_1^2 +
dx_2^2 + dx_3^2)$.  The spatial topology is $T^3$, so $x_i=0$ is identified
with $x_i=1$ for $i=1\ldots 3$.)  For each of these five we sprinkle with a
mean number of elements $\langle N \rangle = 512$.
The sixth is a $4d$ Schwarzschild black hole, where we have used the
technique described in \cite{schwarzschild} to deduce the causal relations.
We sprinkle $\langle N \rangle = 256$ elements into the region $0<t<10$,
$0<r<3M$ in Eddington-Finkelstein coordinates.
Figure \ref{2_3crowns} depicts a similar histogram, for
2-links.
\begin{figure}[htbp]
\begin{center}
\psfrag{M2int}{$\M^2$}
\psfrag{M3int}{$\M^3$}
\psfrag{M4int}{$\M^4$}
\psfrag{M5int}{$\M^5$}
\psfrag{frw}{\hspace{-5mm}FRW}
\includegraphics[width=\textwidth]{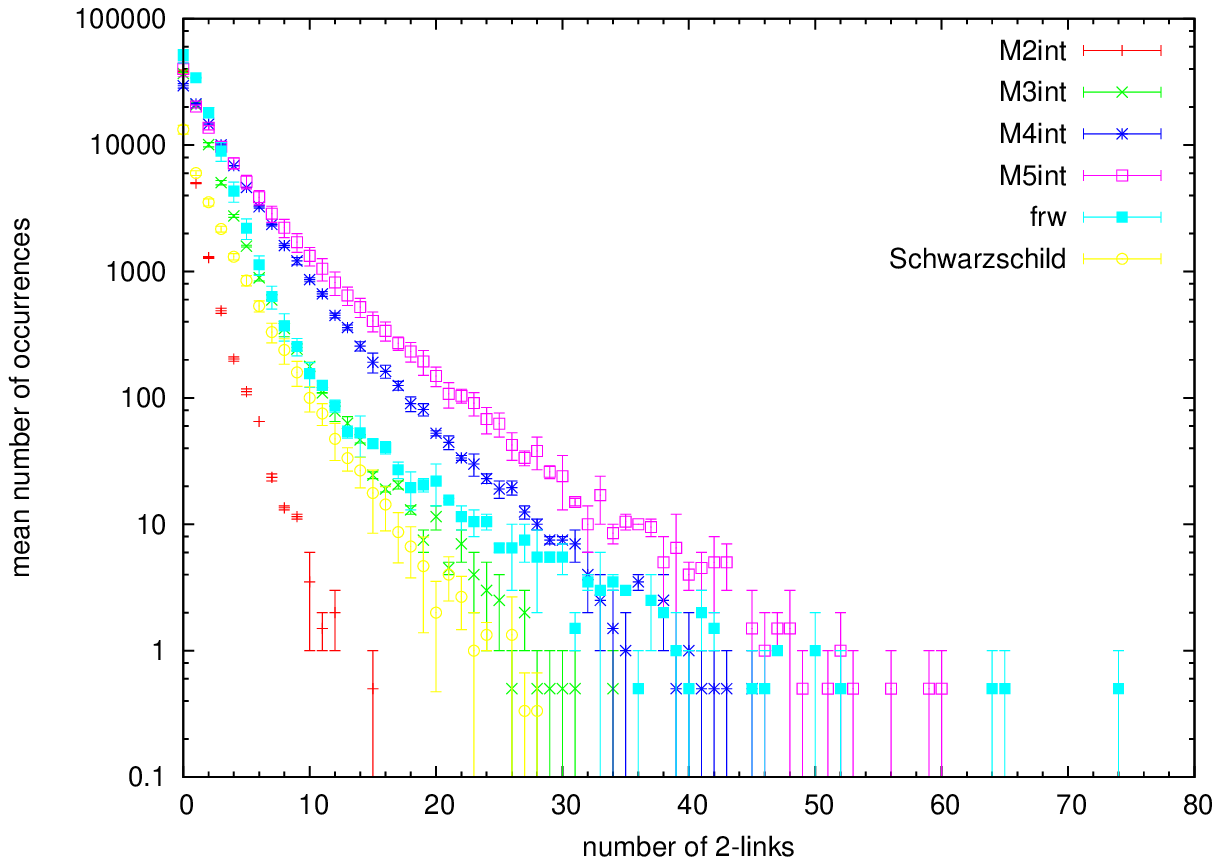}
\end{center}
\caption{2-links in some manifoldlike causal sets.  Each histogram gives the
  mean (and standard error) from two sprinklings.}
\label{manifolds-2l}
\end{figure}
We expect to find an infinite number of 2-links for every 2-antichain, for
spacetime dimension $>2$.  This manifests itself in extremely long
(power law) tails, in
our finite simulations.  Even in two dimensions we find some 2-antichains with
a large number of 2-links, though the fall off is much sharper.
The fact that the results for
the $4d$ Schwarzschild spacetime more closely resemble lower dimensional spacetimes may be
because many of the elements get sprinkled behind the horizon, where the
light cones rapidly fall into the singularity, before they have a chance to
build up a large number of links.

\begin{figure}[htbp]
\begin{center}
\psfrag{M2int}{$\M^2$}
\psfrag{M3int}{$\M^3$}
\psfrag{M4int}{$\M^4$}
\psfrag{cube}{\hspace{-5mm}$1 \!\times\! 1 \!\times\! 1$}
\psfrag{1 1/2 1/4}{$1 \!\times\! \frac{1}{2} \!\times\! \frac{1}{4}$}
\psfrag{1 1/3 1/9}{$1 \!\times\! \frac{1}{3} \!\times\! \frac{1}{9}$}
\psfrag{1 1/4 1/16}{$1 \!\times\! \frac{1}{4} \!\times\! \frac{1}{16}$}
\psfrag{1 1/16 1/256}{$1 \!\times\! \frac{1}{16} \!\times\! \frac{1}{256}$}
\includegraphics[width=\textwidth]{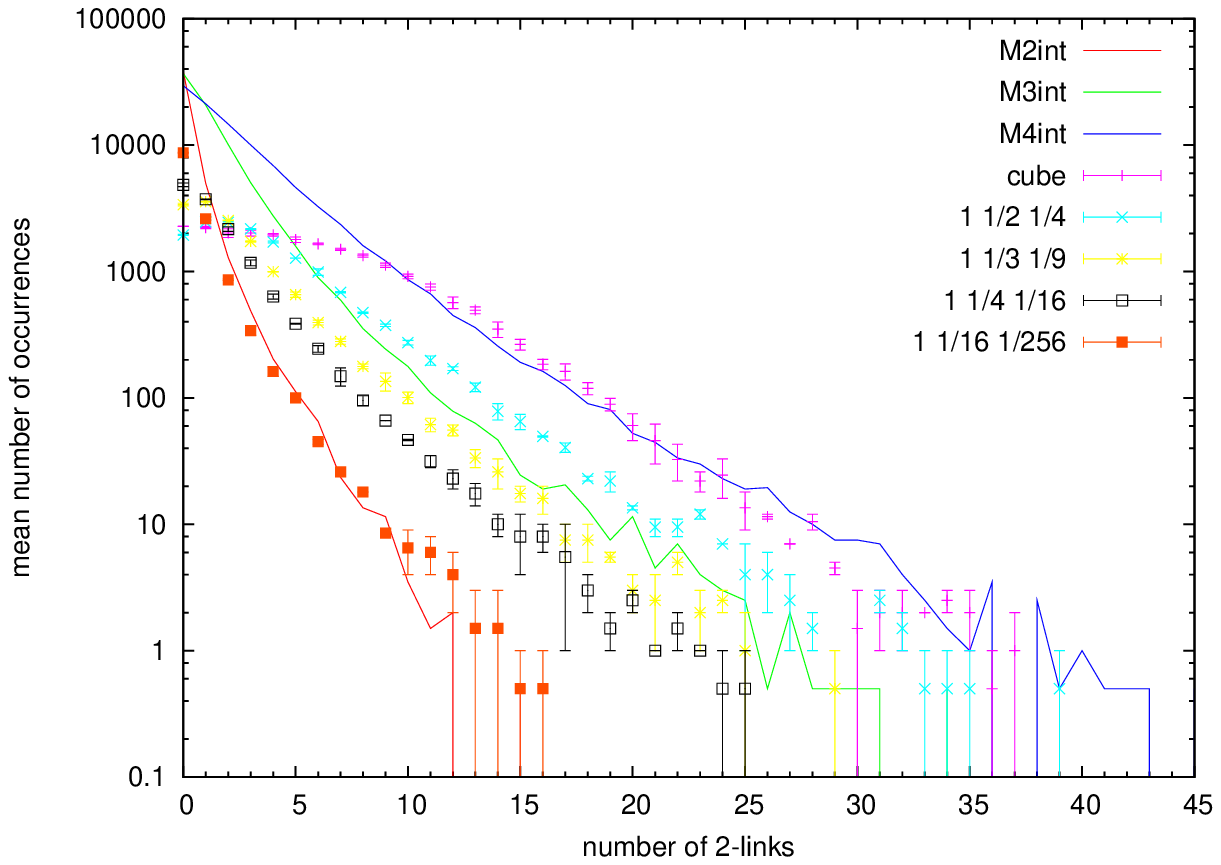}
\end{center}
\caption{2-links in $\langle N \rangle = 256$ sprinklings into $T^3 \times I$
  spacetimes.  The `circumferences' of the toroidal dimensions are given in
  the key.
  Each histogram gives the mean (and standard error) from two sprinklings.  The
  solid lines indicate the means from figure \ref{manifolds-2l}, for comparison.}
\label{kk}
\end{figure}

A popular concept in physics is that we live in some sort of a product
manifold, perhaps with compactified internal dimensions, such as in
Kaluza-Klein or String theory, or with large extra dimensions of a braneworld
scenario.  To investigate how some of these ideas might play out in the causal
set context, we consider sprinklings into a flat spacetime with topology $T^3
\times I$ (i.e.\ a box shaped region of $\M^3$ with opposite sides
identified), in which we vary the size of the dimensions of the torus.  Our
results are given in figure \ref{kk}.
We see that for the case in which all dimensions have the same size (purple),
the results closely match that of the interval in $\M^4$ from figure
\ref{manifolds-2l}.  As the size of the internal dimensions is decreased, the
approximately 256 elements of the causal set begin to be unable to `see' the
internal dimensions, and the asymptotic behaviour of 2-link counts becomes
like those of lower dimensional spacetimes.  In the extreme case of the red
data, with the `internal' dimensions $\leq \frac{1}{16}$ times the size of the
`external', the causal set is not able to resolve the internal dimensions,
and the 2-link counts are almost exactly what one finds for an interval in
$\M^2$.  (As an alternative to what we did here, one could instead hold the
manifold fixed, and sprinkle increasingly more elements to resolve the small
internal dimensions.  We chose this approach of varying the manifold at fixed
$\langle N \rangle$ simply to save on compute time.)

\begin{figure}[htbp]
\begin{center}
\begin{minipage}{7.8cm}
\psfrag{M2}{$\M^2$}
\includegraphics[width=7.8cm]{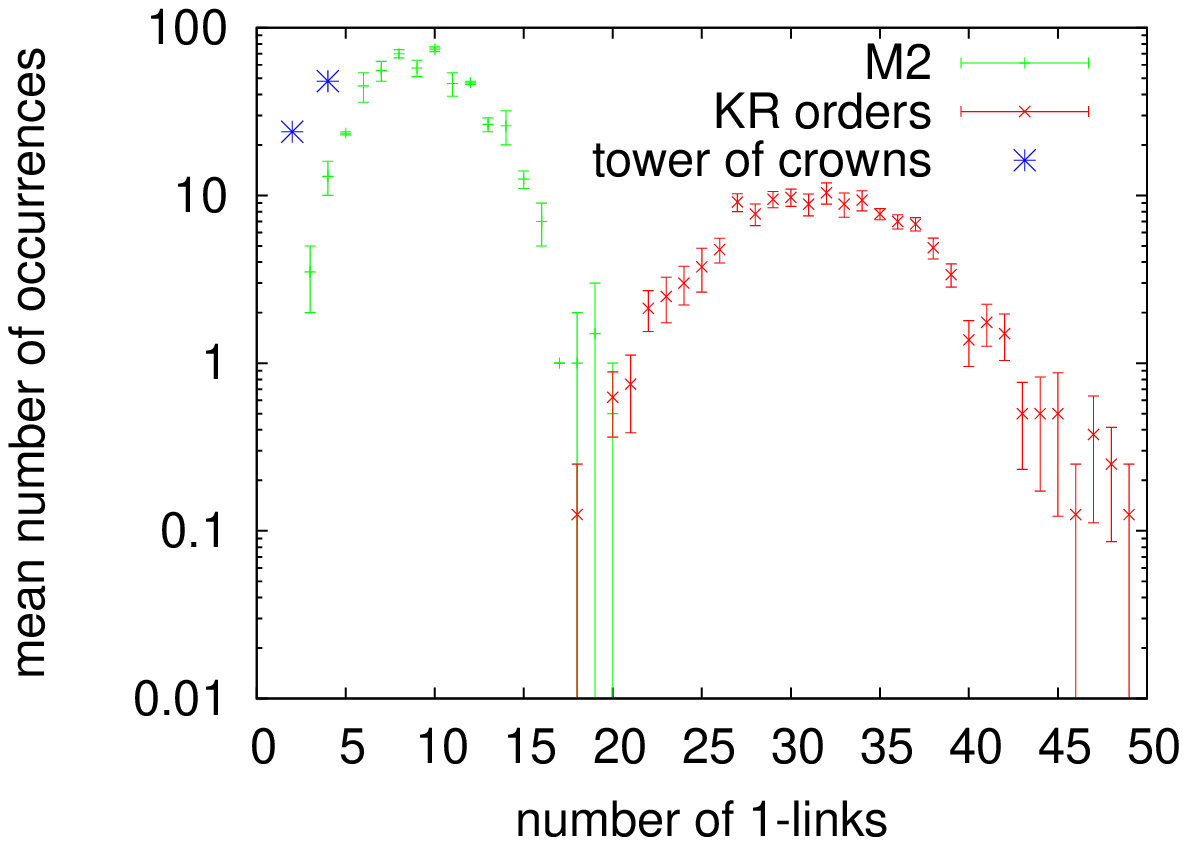}
\caption{\label{nonmanifolds1} 1-links for some non-manifoldlike causal sets:
  (Two) Kleitman-Rothschild order(s) on 64 elements, and a tower of 6
  12-crowns.  Results from sprinkling into $\M^2$ is added for comparison.}
\end{minipage} \hspace{1mm}\begin{minipage}{7.8cm}
\psfrag{M2}{$\M^2$}
\includegraphics[width=7.8cm]{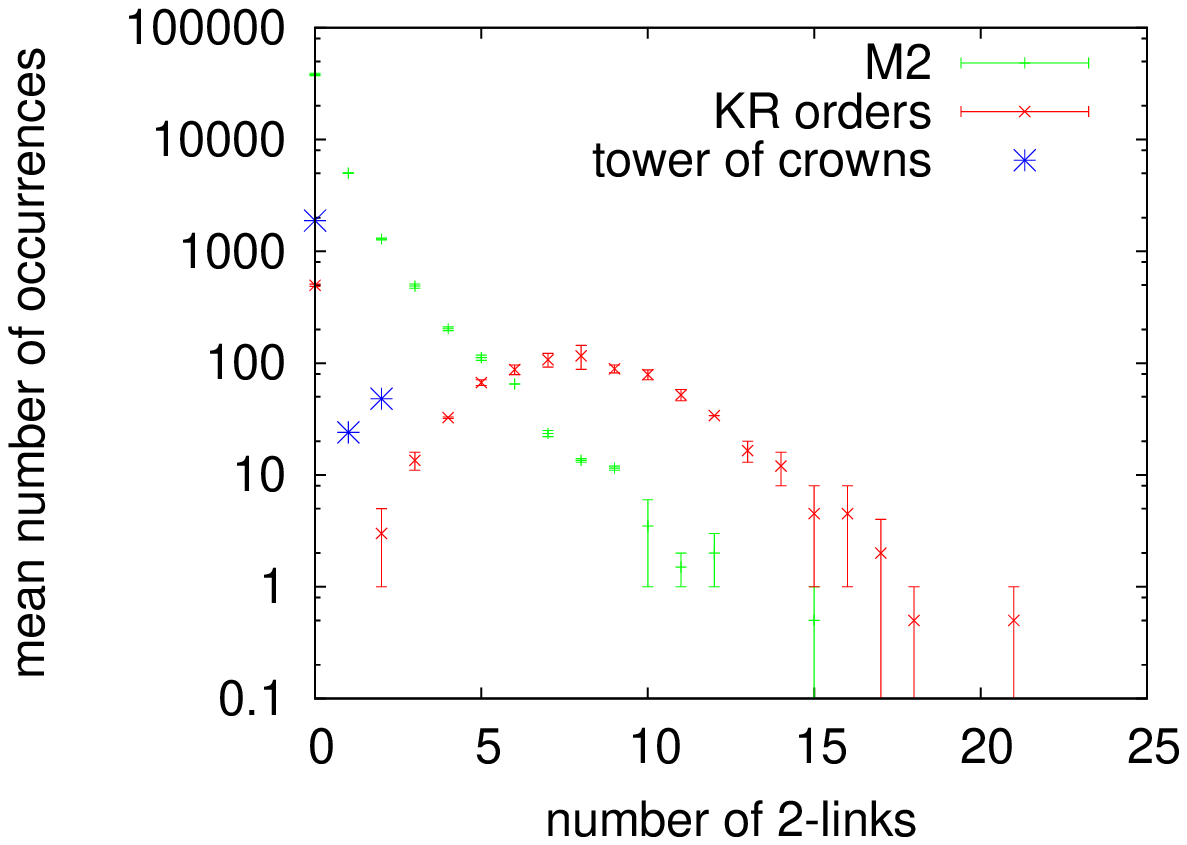}
\caption{\label{nonmanifolds2} 2-links for some non-manifoldlike causal sets:
  (Two) Kleitman-Rothschild order(s) on 64 elements, and a tower of 6
  12-crowns.  Note the large number of pairs in the KR-order (red) without any 2-links.  Results from sprinkling into $\M^2$ is added for comparison.}
\end{minipage} 
\end{center}
\end{figure}

To see how this works as a measure of manifoldlikeness, we try it on two
obviously non-manifoldlike causal sets: the `tower of crowns' and the
generic `Kleitman-Rothschild orders'.
A tower of $m$-crowns consists of a number of layers ($m$-antichains).  If we
label the elements in each layer 
by $0 \ldots m\!-\!1$, then element $i$ in layer $t$ precedes elements $i$ and
$i\!+\!1 \mod m$ in layer $t-1$.
Figure \ref{2_3crowns} depicts a tower of three 3-crowns.
The Kleitman-Rothschild (KR) orders form a generic subset of the set of all finite
partially ordered sets 
\cite{brightwell,kr-orders}.
They consist of three layers, the middle of which contains approximately half
the elements of the causet set, and the top and bottom layer contain about a
quarter each.\footnote{In our simulations we select the cardinality of the
  bottom layer from a Poisson distribution with mean $N/4$ (for an $N$
  element causet set), and likewise for the top layer.  All remaining
  elements are placed in the middle layer.}
For each pair of elements in adjacent layers, we place a relation between
them with probability 1/2.  Every element of the bottom layer precedes every
element of the top layer.
The results for these causal sets are depicted in figures \ref{nonmanifolds1}
and \ref{nonmanifolds2}.
The tower of crowns are obviously not manifoldlike, as the only non-zero bins
are 2 and 4 for the 1-links, and 0, 1, and 2 for the 2-links.  The 1-links histogram
for the KR orders looks like it could have come from a sprinkling.  The
2-links histogram, however, possess a `discontinuity', with a huge spike at
zero followed by empty bins at 1-3.  The spike at zero results from unrelated
pairs of elements, each from different layers.
This sort of behaviour does not resemble sprinklings into spacetime.

\begin{figure}[htbp]
\begin{center}
\psfrag{M2int}{\hspace{-19mm}$\M^2$ interval}
\psfrag{tp15.2}{\hspace{-18mm}$N=2^{15}$, $p=.2$}
\psfrag{B1xI7T55}{\hspace{-13mm}$\M^2$ rectangle}
\psfrag{tp7.7}{\hspace{-20mm}$N=2^{7}$, $p=.7$}
\includegraphics[width=\textwidth]{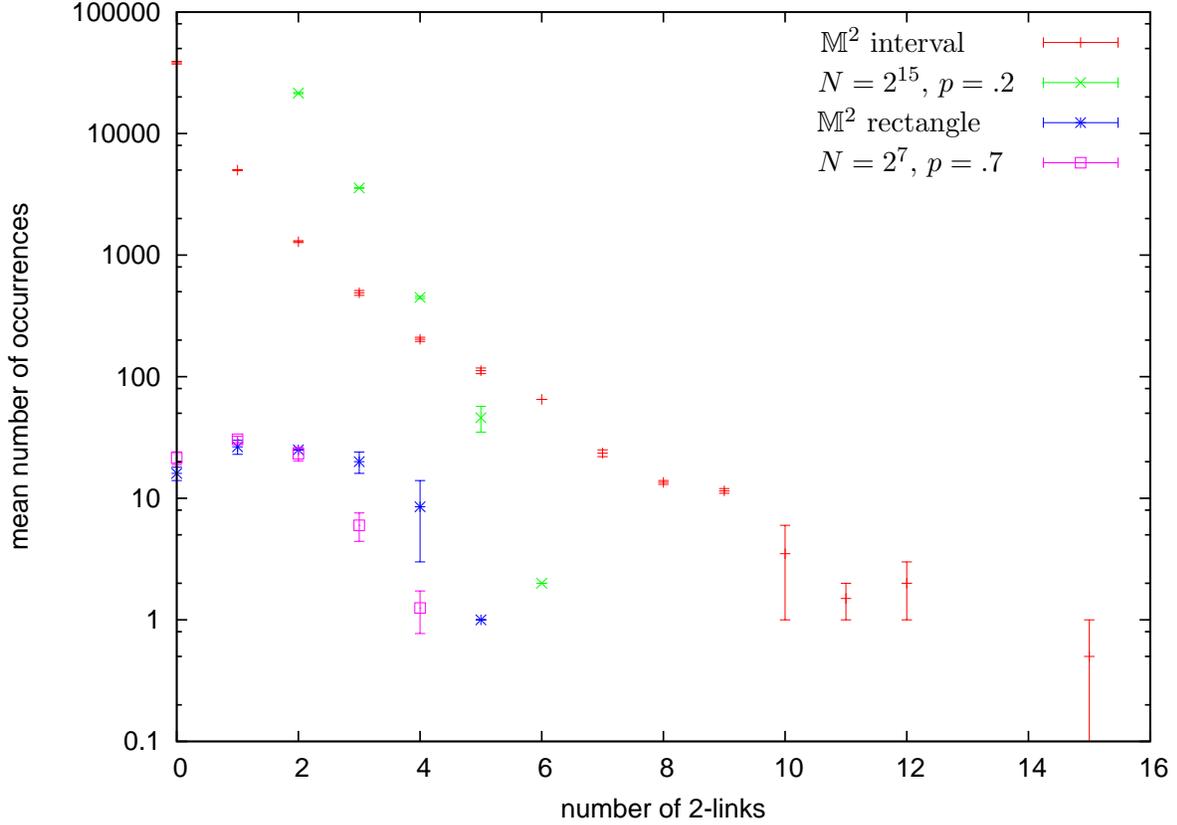}
\end{center}
\caption{Comparison of transitive percolation with sprinklings into $\M^2$.
  The interval in $\M^2$ uses $\langle N \rangle = 512$, while the rectangle
  $\langle N \rangle = 128$.  Each histogram gives the mean and error
  from two causal sets.}
\label{tranperc}
\end{figure}

The last class of causal sets we consider are those generated by the
`transitive percolation' dynamics, which is a simple special case of the generic
class of sequential growth models derived in \cite{cosacc}.  Here one begins
with $N$ elements, and considers every pair of elements $i<j$ in turn,
introducing a relation between them with some fixed probability $p$.  After
relations are introduced in this way, one takes the transitive closure to
arrive at a causal set.  In figure \ref{tranperc} we depict the 2-links for two choices of parameters,
$N=2^{15}$, $p=2/10$, and $N=128$, $p=7/10$.  They fall off very quickly, but
otherwise are not obviously distinct from the results that one gets from
sprinkling into spacetime.  To get some idea as to whether there are
sprinklings into spacetime which mimic these results, we sprinkle into a
rectangle of $\M^2$, with a height/duration 55 times the (spatial) width.
Although the 2-link abundances for transitive percolation at $N=128, p=7/10$
fall off a bit faster than for the sprinkling into the rectangle, it 
remains possible that 
that some choice of parameters for percolation will give a 2-link histogram
which can be matched by a sprinkling.

\section{Conclusion}

A fundamentally atomistic theory of quantum gravity must come with some sort
of relation defined on the discrete elements, in order for nontrivial
geometric structures to emerge.  In the case of causal sets, this relation is
regarded as a causal ordering.  This association with causality makes it
relatively straightforward to see how timelike structures emerge, but spatial
quantities are more subtle.  Here we proposed some quantities, derived from the
discrete causal ordering, which extract spatial information from the causal
set, such as spatial distances.

In order to define spatial distance, we introduced the notion of an $n$-link,
which is a simple derived object which encodes information about the
intersection of light cones of spatially separated elements.  From this we
introduced the 2-link distance, and showed that it is able to overcome the
degeneracy of the former measure of spatial distance for pairs of elements on
a causal set.

In order to recover curved geometry, we further proposed to use the 2-link
distance within some local region, to identify spatial nearest neighbors in
the causal set.  This defines a symmetric, spacelike relation, which is
derived from the causal order.  From it we may hope to recover the length of
continuous curves, and thus derive the metric geometry. 
It is hoped that such spacetime structures will be useful in 
the development of a dynamical law for causal sets.

~\\
The $n$-link captures some element of spatial information, which plain links
(in and of themselves) do not.  In \cite{diameters} is was observed that their
abundance depends somewhat strongly upon dimension, and it was proposed that
they could be used as a dimension estimator, or even as a potentially
stringent condition for a causal set to be faithfully embeddable into a
spacetime manifold.  Here we performed a preliminary investigation of this
`manifoldlikeness' condition, and observed somewhat sharp 
dependence of 2-link abundances on dimension.

In order to see the effect of compact dimensions at various scales, we
explored flat $T^3 \times I$ spacetimes, with varying ratios of scales of the
`internal' dimensions.  We observed that the abundance of 2-links is able to see
the correct dimension of 4 at large sprinkling densities, but sees an
effective, smaller dimension as the sprinkling density is reduced as compared
to the ratio of circumferences of the toroidal dimensions.  The results
mirror those of \cite{meyer_thesis} for embeddings into $S^1 \times I$
spacetime.

The question of how to compute the dimension of discrete spacetime at a range
of length scales is important currently,
because of a number of results from various approaches to quantum gravity
which predict a scale dependent dimension, in particular a smaller fractal
dimension at high energies \cite{leonardo_dimension} (and references therein).
In addition to the older methods
mentioned in the introduction,
counting of $n$-links may provide an important alternate method to deduce which
dimensions arise dynamically at various scales in causal set quantum
gravity.

~\\
In section \ref{manifoldlikeness} we presented some preliminary results on a
manifoldness test based on counting $n$-links.  It is easily able to
distinguish some obviously non-manifoldlike causal sets from those which are
faithfully embeddable into
spacetime, however it
had some difficulty in distinguishing causal sets
generated from transitive percolation from those of sprinklings into regions
of Minkowski space.  This may be rectified when considering $n$-links for $n>2$,
or by performing a careful analysis on the parameter space of the
transitive percolation and sprinkling models.  
Of course it can only be a necessary condition for there to exist a faithful
embedding into spacetime, and not sufficient, because one can always compose
partial orders with a particular form for the $n$-link counts, which is not
manifoldlike.

\ack
This research was supported primarily by the Perimeter Institute for
Theoretical Physics.  Research at Perimeter Institute is supported by the
Government of Canada through Industry Canada and by the Province of Ontario
through the Ministry of Research \& Innovation.  Earlier work was supported by
the Marie Curie Research and Training Network ENRAGE (MRTN-CT-2004-005616),
and the Royal Society International Joint Project 2006-R2.
Some numerical results were made possible by the facilities of the
Shared Hierarchical Academic Research Computing Network
(SHARCNET:www.sharcnet.ca).

\section*{References}
\providecommand{\newblock}{}

\end{document}